\documentclass[pre, aps, showpacs, showkeys]{revtex4}
\usepackage{graphicx}
\begin{document}

\title{Intrinsic localized modes in dust lattices
\footnote{Proceedings of the \textit{International Conference on
Plasma Physics - ICPP 2004}, Nice (France), 25 - 29 Oct. 2004;
contribution P1-104; available online at:
\texttt{http://hal.ccsd.cnrs.fr/ccsd-00001892/en/} .}}

\author{Ioannis Kourakis$^{1, }$\footnote{On leave from: U.L.B. -
Universit\'e Libre de Bruxelles, Physique Statistique et Plasmas
C. P. 231, Boulevard du Triomphe, B-1050 Brussels, Belgium; also:
Facult\'e des Sciences Apliqu\'ees - C.P. 165/81 Physique
G\'en\'erale, Avenue F. D. Roosevelt 49, B-1050 Brussels, Belgium;
\\Electronic address: \texttt{ioannis@tp4.rub.de}},
Vassileios Basios$^{2, }$\footnote{Electronic address:
\texttt{vbasios@ulb.ac.be}} and Padma Kant Shukla$^{1,
}$\footnote{Electronic address: \texttt{ps@tp4.rub.de}}}
\affiliation{$^{1}$ Institut f\"ur Theoretische Physik IV,
Fakult\"at f\"ur Physik und Astronomie,
Ruhr--Universit\"at Bochum, D-44780 Bochum, Germany \\
$^{2}$ Universit\'e Libre de Bruxelles, Centre for Nonlinear
Phenomena and Complex Systems, C.P. 231 Physique Chimique,
Boulevard du Triomphe, B-1050 Brussels, Belgium }
\date{\today}

\begin{abstract}
Intrinsic Localized Modes (ILM) (or Discrete Breathers, DB) are
localized oscillatory modes known to occur in atomic or molecular
chains characterized by coupling and/or on-site potential
nonlinearity. Quasi-crystals of charged mesoscopic dust grains
(dust lattices), which have been observed since hardly a decade
ago, are an exciting paradigm of such a nonlinear chain. In
gas-discharge experiments, these crystals are subject to forces
due to an externally imposed electric and/or magnetic field(s),
which balance(s) gravity at the levitated equilibrium position, as
well as to electrostatic inter-grain interaction forces. Despite
the profound role of nonlinearity, which may be due to inter-grain
coupling, mode-coupling and to the sheath environment, the
elucidation of the nonlinear mechanisms governing dust crystals is
still in a preliminary stage. This study is devoted to an
investigation, from very first principles, of the existence of
discrete localized modes in dust layers. Relying on a set of
evolution equation for transverse
charged grain displacements, we examine the conditions for the
existence and sustainance of discrete localized modes and  discuss
the dependence of their characteristics on intrinsic plasma
parameters. In addition, the possibility of DB stabilisation via
an external force is discussed.
\end{abstract}
option if keyword

\maketitle

\section{Introduction}

A variety of linear and nonlinear collective effects are known to
occur in a dust-contaminated plasma (\textit{dusty plasma}, DP)
\cite{psbook} and relative theoretical research has received new
impulse, since roughly a decade ago, thanks to laboratory and
space dusty plasma observations. An issue of particular importance
in DP research is the formation of strongly coupled DP crystals by
highly charged dust grains, typically in the sheath region above a
horizontal negatively biased electrode in experiments
\cite{psbook, Morfill}.
Low-frequency oscillations are known to occur \cite{Morfill} in
these mesoscopic dust grain quasi-lattices in the longitudinal
($\sim \hat x$, in-plane, acoustic mode), horizontal transverse
($\sim \hat y$, in-plane, shear mode) and vertical transverse
($\sim \hat z$, off-plane, optic-like mode) directions.

Various types of localized (nonlinear) excitations are known from
solid state physics to exist in periodic chains
(\textit{lattices})of interacting particles, in addition to
propagating vibrations (\textit{phonons}), due to a mutual balance
between the intrinsic nonlinearity of the medium and dispersion.
Such structures, usually investigated in a continuum approximation
(i.e. assuming that the typical spatial variation scale far
exceeds the typical lattice scale, e.g. the lattice constant
$r_0$), include non-topological \emph{solitons} (pulses),
\emph{kinks} (i.e. shocks or dislocations) and localized modulated
envelope structures (\emph{envelope solitons}), and generic
nonlinear theories have been developed in order to investigate
their relevance in different physical contexts \cite{Remo}. In
addition to these (continuum) theories, which deliberately
sacrifice discreteness in the altar of analytical tractability,
attention has been paid since more than a decade ago to highly
localized (either stationary or propagating) vibrating structures
[e.g. \emph{discrete breathers} (DBs), also widely referred to as
\textit{intrinsic localized modes} (ILMs)], which owe their very
existence to the lattice discreteness itself. Thanks to a few
pioneering works \cite{Page, Dauxois, Kivshar, McKay1, Flach1} and
a number of studies which followed, many aspects involved in the
spontaneous formation, mobility and interaction of DBs are now
elucidated, both theoretically and experimentally; see in Refs.
\cite{Flach2, Chaos, Campbell} for a review (also see Refs.
\cite{Jeroen, Bountis1}, with reference to this study).

Despite the fact that nonlinearity is an inherent feature of the
dust crystal dynamics (either due to inter-grain electrostatic
interactions, to the sheath environment, which is intrinsically
anharmonic, or to coupling between different degrees of freedom),
our knowledge of nonlinear mechanisms related to dust lattice
modes still appears to be in a rather preliminary stage today.
Small amplitude localized longitudinal excitations (described by a
Boussinesq equation for the longitudinal grain displacement $u$,
or a Korteweg-deVries equation for the density $\partial
u/\partial x$) were considered in Refs. \cite{Melandso} and
generalized in Ref. \cite{IKPKSEPJDsols}. The nonlinear amplitude
modulation of longitudinal and transverse (vertical, off-plane)
dust lattice waves was recently considered in Refs. \cite{AMS2,
IKPKSLDLWMI}  and \cite{IKPKSTDLWMI, IKPKSTMDLWMI} (also see
\cite{Ivlev-Zhdanov}), respectively. In fact, all of these studies
rely on a \emph{continuum} description of the dust lattice. On the
other hand, the effect of the high discreteness of dust crystals,
clearly suggested by experiments \cite{Ivlev2000, Misawa, Zafiu,
Liu}, may play an important role in mechanisms like energy
localization, storage and propagation and thus modify the
crystal's dynamical response to external excitations (in view of
DP application design, e.g.). To the very best of our knowledge,
no study has been carried out, from first principles, of the
relevance of DB excitations with respect to dust lattice waves,
apart from a preliminary investigation (restricted to single-mode
transverse dust-breathers), which was recently presented
\cite{IKPKSPOPDB}. This text aims in making a first analytical
step towards filling this gap, by raising a number of questions
which have not been addressed before. This study is neither
exhaustive nor complete; it will be complemented by forthcoming
work.

\section{The model}

We shall consider the vertical (off-plane, $\sim \hat z$) grain
displacement in a dust crystal (assumed quasi-one-dimensional:
identical grains of charge $q$ and mass $M$ are situated at $x_n =
n\, r_0 , \,$\ where $n= ..., -1, 0, 1, 2, ...$), by taking into
account the intrinsic nonlinearity of the sheath electric (and/or
magnetic) potential. The in-plane (longitudinal, acoustic, $\sim
\hat x$ and shear, optical, $\sim \hat y$) degrees of freedom are
assumed suppressed; this situation is indeed today realized in
appropriate experiments, where an electric potential (via a thin
wire) \cite{Ivlev2000} or a coherent light (laser) impulse
\cite{Misawa, Zafiu, Liu} is used to trigger transverse dust grain
oscillations, while (a) confinement potential(s) ensure(s) the
chain's in-plane stability.

\subsection{Equation of motion}

The vertical grain displacement obeys an equation in the form
\cite{IKPKSTDLWMI, IKPKSTMDLWMI}
\begin{equation}
\frac{d^2 \delta z_n}{dt^2} + \nu \frac{d \delta z_n}{dt} + \,
\omega_0^2 \, (\,\delta z_{n+1} + \,\delta z_{n-1} - 2 \,\delta
z_n) + \omega_g^2 \, \delta z_n
+ \alpha \, (\delta z_n)^2 + \beta \, (\delta z_n)^3  = 0 \, ,
\label{eqmotion}
\end{equation}
where $\delta z_n(t) = z_n(t) - z_0$ denotes the small
displacement of the $n-$th grain around the (levitated)
equilibrium position $z_0$, in the transverse ($z-$) direction.
The characteristic frequency $\omega_0\,  = \bigl[ - q
\Phi'(r_0)/(M r_0) \bigr]^{1/2}$ results from the dust grain
(electrostatic) interaction potential  $\Phi(r)$, e.g. for a
Debye-H\"uckel potential  \cite{Konopka}: \( \Phi_D(r) = ({q}/{r})
\,e^{-{r/\lambda_D}}\), one has: \( \omega_{0, D}^2\,  = q^2/(M
r_0^3) \, (1 + r_0/\lambda_D) \,\exp(-r_0/\lambda_D) \, ,
\) where $\lambda_D$ denotes the effective DP Debye radius
\cite{psbook}. The damping coefficient $\nu$ accounts for
dissipation due to collisions between dust grains and neutral
atoms. The gap frequency $\omega_g$ and the nonlinearity
coefficients $\alpha, \beta$ are defined via the overall vertical
force:
\begin{equation}
F(z) = F_{e/m} - Mg \approx - M [\omega_g^2 \delta z_n + \alpha \,
(\delta z_n)^2 + \beta \, (\delta z_n)^3 ] \, + {\cal O}[(\delta
z_n)^4] \, , \end{equation} which has been expanded around $z_0$ 
by formally taking into account the (anharmonicity of the) local
form of the sheath electric (follow exactly the definitions in
Ref. \cite{IKPKSTDLWMI}, not reproduced here) and/or magnetic
\cite{comment1} field(s), as well as, possibly, grain charge
variation due to charging processes \cite{IKPKSTMDLWMI}. Recall
that the electric/magnetic levitating force(s) $F_{e/m}$
balance(s) gravity at $z_0$. Notice the difference in structure
from the usual nonlinear Klein-Gordon equation used to describe
one-dimensional oscillator chains --- cf. e.g. Eq. (1) in Ref.
\cite{Kivshar}: TDLWs (\textit{`phonons'}) in this chain are
stable only in the presence of the field force $F_{e/m}$.

For convenience, we may re-scale the time and vertical
displacement variables over appropriate quantities, i.e. the
characteristic (single grain) oscillation period $\omega_g^{-1}$
and the lattice constant $r_0$, respectively, viz. $t =
\omega_g^{-1} \tau$ and $\delta z_n = r_0 q_n$; Eq.
(\ref{eqmotion}) is thus expressed as:
\begin{equation}
\frac{d^2 q_n}{d \tau^2} + \, \epsilon (\,q_{n+1} + \, q_{n-1} - 2
\,q_n) + \, q_n + \alpha' \, q_n^2 + \beta' \, q_n^3 = 0 \, ,
\label{eqmotion1}
\end{equation}
where the (dimensionless) damping term, now expressed as
$({\nu}/{\omega_g}) {d q_n}/{d\tau} \equiv \nu' \dot{q}_n$, will
be henceforth omitted in the left-hand side. The coupling
parameter $\epsilon = {\omega_0^2}/{\omega_g^2}$ measures the
strength of the inter-grain interactions (with respect to the
single-grain vertical vibrations); this is typically a
\emph{small} parameter, in real experiments (see below). The
nonlinearity coefficients are now: $\alpha' = \alpha
r_0/\omega_g^2$ and $\beta' = \beta r_0^2/\omega_g^2$.

Eq. (\ref{eqmotion1}) will be the basis of the analysis that will
follow. Note that the primes in $\alpha'$ and $\beta'$ will
henceforth be omitted.

\subsection{The model Hamiltonian}

In order to relate our physical problem to existing generic models
from solid state physics, it is appropriate to consider the
equation of motion (\ref{eqmotion}) as it may be derived from a
Hamiltonian function, which here reads:
\begin{equation}
H = \sum_{j = 1}^N \biggl[\frac{p_j^2}{2 m_j} \, + \, V(q_j) \, -
\, \frac{\epsilon}{2} (q_j - q_j-1)^2 \biggr] \, .
\label{hamiltonian}
\end{equation}
Here, $p_j$ obviously denotes the (classical) momentum $p_j = M
\dot{q}_j$. The substrate potential, related to the sheath plasma
environment, is of the form:
\begin{equation}
V(q_j) \, = \, \frac{1}{2} \,q_j^2\, +  \, \frac{\alpha}{3}
\,q_j^3 \,+\, \frac{\beta}{4}\, q_j^4 \, . \label{substrate}
\end{equation}
The coupling parameter $\epsilon$ takes \emph{small} numerical
values (cf. below), accounting for the high lattice discreteness
anticipated in this study. The minus sign preceding it denotes the
inverse dispersive character of (linear excitations propagating
in) the system; see the discussion below. Upon setting $\epsilon
\rightarrow -\epsilon$, the `traditional' (discretized)
\textit{nonlinear Klein-Gordon} model is recovered
\cite{commentTB}.

It should be noted that both experimental \cite{Ivlev2000} and ab
initio (numerical) \cite{Sorasio} studies suggest that dust
crystals are embedded in an \emph{nonlinear} on-site (sheath)
potential $V$, in the vertical direction, which is (possibly
strongly) asymmetric around the origin, i.e. \emph{not} an
\emph{even} function of $q_j$. This implies a finite value of the
cubic anharmonicity parameter $\alpha$, thus invalidating models
involving \emph{even} potential forms -- e.g. $V(q_j) \sim q_j^2/2
+ \beta q_j^4/4$ -- in our case.

\section{Linear waves}

Retaining only the linear contribution and considering
oscillations of the type, $\delta z_n \sim \,\exp[i \,(k n r_0 -
\omega t)] + c.c.$ (complex conjuguate) in Eq. (\ref{eqmotion}),
one obtains the well known transverse dust lattice (TDL) wave
optical-mode-like dispersion relation
\begin{equation}
\omega^2\,  = \omega_g^2\, - 4 \omega_0^2\, \sin^2 \biggl( \frac{k
r_0}{2} \biggr) \, , \label{dispersion-discrete}
\end{equation}
i.e.
\begin{equation}
\tilde \omega^2 = 1\, - 4 \epsilon \, \sin^2 ({\tilde k}/{2}) \, .
\label{dispersion-discrete1}
\end{equation}
\begin{figure}[htb]
 \centering
 \resizebox{6.5cm}{!}{
 \includegraphics[]{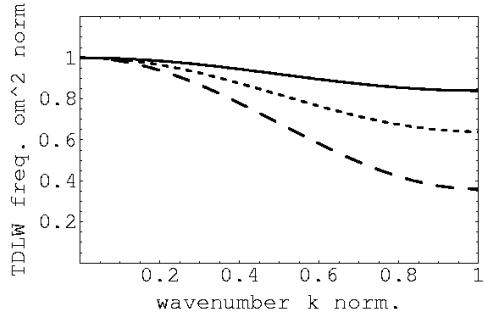}
} \caption{\small{The dispersion relation of the TDL excitations:
frequency $\omega$ (normalized over $\omega_g$) versus wavenumber
$k$. The value of $\omega_0/\omega_g$ ($\sim$ coupling strength)
increase from top to bottom. Note that upper (less steep,
continuous) curve is more likely to occur in a real
(weakly-coupled) DP crystal.}} \label{fig1}
\end{figure}
See that the wave frequency $\omega \equiv \tilde \omega \omega_g$
\emph{decreases} with increasing wavenumber $k = 2 \pi/\lambda
\equiv \tilde k/r_0$ (or decreasing wavelength $\lambda$),
implying that transverse vibrations propagate as a \emph{backward
wave}: the group velocity $v_g = \omega'(k)$ and the phase
velocity $\omega_{ph}=\omega/k$ have opposite directions (this
behaviour has been observed in recent experiments). The
modulational stability profile of these linear waves (depending on
the plasma parameters) was investigated in Refs.
\cite{IKPKSTDLWMI, IKPKSTMDLWMI}. Notice the natural \emph{gap
frequency} $\omega(k=0)\, = \omega_g = \omega_{max}$,
corresponding to an overall motion of the chain's center of mass,
as well as the \emph{cutoff frequency} $\omega_{min}\, =
(\omega_g^2\, - 4 \omega_0^2)^{1/2} \equiv \omega_g\, (1 - 4
\epsilon^2)^{1/2}$ (obtained at the end of the first Brillouin
zone $k = \pi/r_0$) which is \emph{absent in the continuum limit},
viz. $\omega^2\, \approx \omega_g^2\, - \omega_0^2 \, k^2 \,
r_0^2$ (for $k \ll r_0^{-1}$); obviously, the study of wave
propagation in this ($k \lesssim \pi/r_0$) region invalidates the
continuum treatment employed so far in literature. The essential
feature of discrete dynamics, to be retained here, is the (narrow)
bounded TDLW (\textit{`phonon'}) frequency band, limited in the
interval $\omega \in [(\omega_g^2\, - 4 \omega_0^2)^{1/2},
\omega_g]$; note that one thus naturally obtains  the stability
constraint: $\omega_0^2/\omega_g^2 = \epsilon < 1/4$ (so that
$\omega \in \Re \quad \forall k \in [0, \pi/r_0]$).

 We needn't go into further details concerning the
linear regime, since it is covered in the literature. We shall,
instead, see what happens if the {\em nonlinear} terms are
retained, in this discrete description.

\section{Existence of discrete breathers - analysis}

We are interested in the (possibility for the) existence of
multi-mode breathers, i.e. localized (discrete) excitations in the
form:
\begin{equation}
q_n(\tau) = \sum_{m=-\infty}^\infty A_n(m) \exp(i m \omega \tau)
\, , \label{DB-form}
\end{equation}
with $A_n(m) = A_n^*(-m)$ for reality and $|A_n(m)| \rightarrow 0$
as $n \rightarrow \pm \infty$, for localization.

\subsection{The formalism}

Inserting Eq. (\ref{DB-form}) in the equation of motion
(\ref{eqmotion1}), one obtains a (numerable) set of algebraic
equations in the form:
\begin{eqnarray}
A_{n+1}(m) + A_{n-1}(m) + C_m A_{n}(m) \, &=& \,-
\frac{\beta}{\epsilon} \sum_{m_1}\sum_{m_2}\sum_{m_3} A_n(m_1)
A_n(m_2) A_n(m_3) \nonumber \\
& & - \,\frac{\alpha}{\epsilon} \sum_{m_4}\sum_{m_5} A_n(m_4)
A_n(m_5) \, , \label{equations-multi}
\end{eqnarray}
where the dummy indices $m_j$ ($j = 1, 2, ..., 5$) satisfy $m_1 +
m_2 + m_3 = m_4 + m_5 = m$; we have defined:
\begin{equation} C_m = - \biggl( 2 - \frac{1 - m^2 \omega^2}{\epsilon}
\biggr) \, . \label{Cm}
\end{equation}

In order to be more precise and gain in analytical tractability
(yet somewhat losing in generality), one may assume that the
contribution of higher (for $m \ge 2$) frequency harmonics may be
neglected. Eq. (\ref{DB-form}) then reduces to:
\begin{equation}
q_n(t) \approx 2 A_n(1) \cos \omega \tau \, + A_n(0) \, .
\label{DB-form-approx}
\end{equation}
Note the zeroth-harmonic (mean displacement) term, for $n=0$,
which is due to the cubic term ($\sim \alpha$, above), and should
vanish for $\alpha = 0$. The system (\ref{equations-multi}) thus
becomes (for $m = 0, 1$):
\begin{eqnarray} A_{n+1}(1) + A_{n-1}(1) + C_1 A_n(1) \, &=& \, - 2
\frac{\alpha}{\epsilon} A_n(1) A_n(0)  - \frac{\beta}{\epsilon}
 \, [A_n(1) A_n^2(0) + 3 A_n^2(1) A_n(-1)]
\nonumber \\
A_{n+1}(0) + A_{n-1}(0) + C_0 A_n(0) \, &=& \, - 2
\frac{\alpha}{\epsilon} A_n(1) A_n(-1) \, - 6
\frac{\beta}{\epsilon} \, A_n(0) A_n(1) A_n(-1) \, ,
\label{equations-single}
\end{eqnarray}
i.e., setting $A_n(1) = A_n(-1) = A_n$ and $A_n(0) = B_n$, viz.
$q_n(t) = 2 A_n \cos \omega \tau \, + B_n$:
\begin{eqnarray} A_{n+1} + A_{n-1} + C_1 A_n \, &=& \, - 2
\frac{\alpha}{\epsilon} A_n B_n  -  \frac{\beta}{\epsilon}
 \, (A_n B_n^2 + 3 A_n^3)
\nonumber \\
B_{n+1} + B_{n-1} + C_0 B_n \, &=& \, - 2 \frac{\alpha}{\epsilon}
A_n^2 \, - 6 \frac{\beta}{\epsilon} \, A_n^2 B_n \, .
\label{equations-single1}
\end{eqnarray}
We see that the amplitudes $A_n$ ($B_n$) of the first (zeroth)
harmonic terms, corresponding to the $n-$th site, will be given by
the iterative solution of Eqs. (\ref{equations-single1}) [or, of
Eqs. (\ref{equations-multi}), should higher harmonics $m$ be
considered]. In  specific, one may express
(\ref{equations-single1}) as:
\begin{eqnarray} a_{n+1} &=& - c_n - C_1 a_n + 2
\frac{\alpha}{\epsilon} a_n b_n  + \frac{\beta}{\epsilon}
 \, (a_n b_n^2 + 3 a_n^3) \equiv f_1(a_n, b_n, c_n, d_n)
\nonumber \\
b_{n+1} &=& - d_n - C_0 b_n + 2 \frac{\alpha}{\epsilon} a_n^2 \, +
6 \frac{\beta}{\epsilon} \, a_n^2 b_n
 \equiv f_0(a_n, b_n, c_n, d_n) \nonumber \\
c_{n+1} &=& a_n \nonumber \\d_{n+1} &=& b_n \, ,
\label{equations-single-map}
\end{eqnarray}
and then iterate, for a given initial condition $(a_1, b_1, c_1,
d_1) = (A_1, B_1, A_0, B_0)$, the map defined by
(\ref{equations-single-map}).

 At this stage,
one needs to determine whether the fixed point of the
4-dimensional map (\ref{equations-single-map}) [or of the complete
4N-dimensional map corresponding to (\ref{equations-multi}), in
general] is hyperbolic, and examine the dimensionality of its
stable and unstable manifolds. It is known \cite{Jeroen, Bountis1}
that the existence of discrete breathers is associated with
homoclinic orbits, implying a saddle point at the origin.

Let us now linearize the map (\ref{equations-single-map}) near the
fixed point $(a_1, b_1, c_1, d_1) = (0, 0, 0, 0) \equiv
\mathbf{0}_4$, by setting e.g. $(a_n, b_n, c_n, d_n) = (\xi_1,
\xi_2, \xi_3, \xi_4)_{n}^T \equiv \mathbf{\Xi}_n \in \Re^4$, where
$\xi_{j, n} \ll 1$ ($j=1, ..., 4$). One thus obtains the matrix
relation:
\begin{equation}
\mathbf{\Xi}_{n+1} = \mathbf{M} \, \mathbf{\Xi}_n \, ,
\end{equation}
where $\mathbf{M}$ is the matrix:
\begin{equation}
\mathbf{M} = \left(
\begin{array}{cccc}
- C_1 & 0 & -1 &  0 \\
0 & - C_0 & 0 &  -1 \\
1 & 0 & 0 &  0 \\
0 & 1 & 0 &  0 \\
\end{array}
\right) \, .
\end{equation}
Now, it is a trivial algebraic exercise to show that the
characteristic polynomial $p(\lambda) \equiv Det(\mathbf{M} -
\lambda \mathbf{I})$ of this matrix may be factorized as:
\[
p(\lambda)\,=\,(\lambda^2 + C_0 \lambda +1)\, (\lambda^2 + C_1
\lambda +1) \equiv p_0(\lambda) p_1(\lambda) \, ,
\]
implying the existence of 4 eigenvalues, say $\lambda_{1, 2, 3,
4}$, such that $p_0(\lambda_{1, 2}) = p_0(\lambda_{3, 4}) = 0$.
One may check that the condition for all eigenvalues to be real
and different, hence for $\mathbf{0}_4$ to be a saddle point,
amounts to the constraint: $|C_{0, 1}| > 2$, i.e. \( C_{0} \notin
[-2, 2] \) \emph{and} \( C_{1} \notin [-2, 2] \). Recalling that
\begin{equation} C_1 = (1 - 2 \epsilon - \omega^2)/\epsilon \, ,
\qquad C_0 = (1 - 2 \epsilon)/\epsilon , \label{C01}
\end{equation}
from (\ref{Cm}), one finds the (simultaneous) constraints: $1 - 4
\epsilon > 0$ and $(1 - \omega^2) (1 - \omega^2 - 4 \epsilon) >
0$. One immediately sees that the former (i.e. $\epsilon < 1/4$)
corresponds to the linear stability condition mentioned above,
while the latter amounts to the requirement that the breather
frequency should lie outside the `phonon band', viz.
$\omega^2/\omega_g^2 \notin [1-4 \epsilon, 1]$.

|

It is straightforward to show that in case one considers the
complete multi-mode map, defined by Eq. (\ref{equations-multi}),
one obtains an analogous factorizable characteristic polynomial
for the $4N\times4N$ matrix $\mathbf{M}$, viz. $p(\lambda) =
\prod_m p_m(\lambda)$. The same analysis then leads to the
hyperbolicity criterion:
\[
|C_m| < 2 \qquad m = 0, 1, 2, ...\] One thus recovers, in addition
to the first of the above constraint ($\epsilon < 1/4$), the
condition: $m \omega/\omega_g \notin  (1-4 \epsilon, 1)^{1/2}$
($\forall m = 0, 1, 2, ...$), which coincides with the --
physically meaningful -- non-breather-phonon-resonance condition
found via different analytical methods \cite{Flach1, Flach2,
Chaos}. We see that the breather frequency, as well as all its
multiples (harmonics) should lie outside the allowed linear
vibration frequency band, otherwise the breather may enter in
resonance with the linear TDLW (\textit{`phonon'}) dispersion
curve, resulting in its being decomposed into a superposition of
linear excitations (and hence de-localized).

\subsection{Numerical analysis}

At this stage, one is left with task of finding the numerical
values of $A_n, B_n$ [cf. (\ref{equations-single1})] for a given
homoclinic orbit; these may then be used as an initial condition,
in order to solve the equation (\ref{equations-single1})
numerically,  by considering a given number of particles $N$ and
harmonic modes $m_{max}$ (viz. $m = 0, 1, 2, ..., m_{max}$). One
thus obtains a given set of numerical values for $u_n$ ($n=1, 2,
..., N$), which constitute the numerical solution for the
anticipated breather excitation. The stability of the solution
thus obtained, say $\hat q_n$, my be checked by  directly
substituting with $q_n = \hat q_n + \xi_n$ (for $n = -N, ..., 0,
..., N$) into the initial equation of motion (\ref{eqmotion1}).

\begin{figure}[htb]
 \centering
 \resizebox{15.5cm}{!}{
 \includegraphics[]{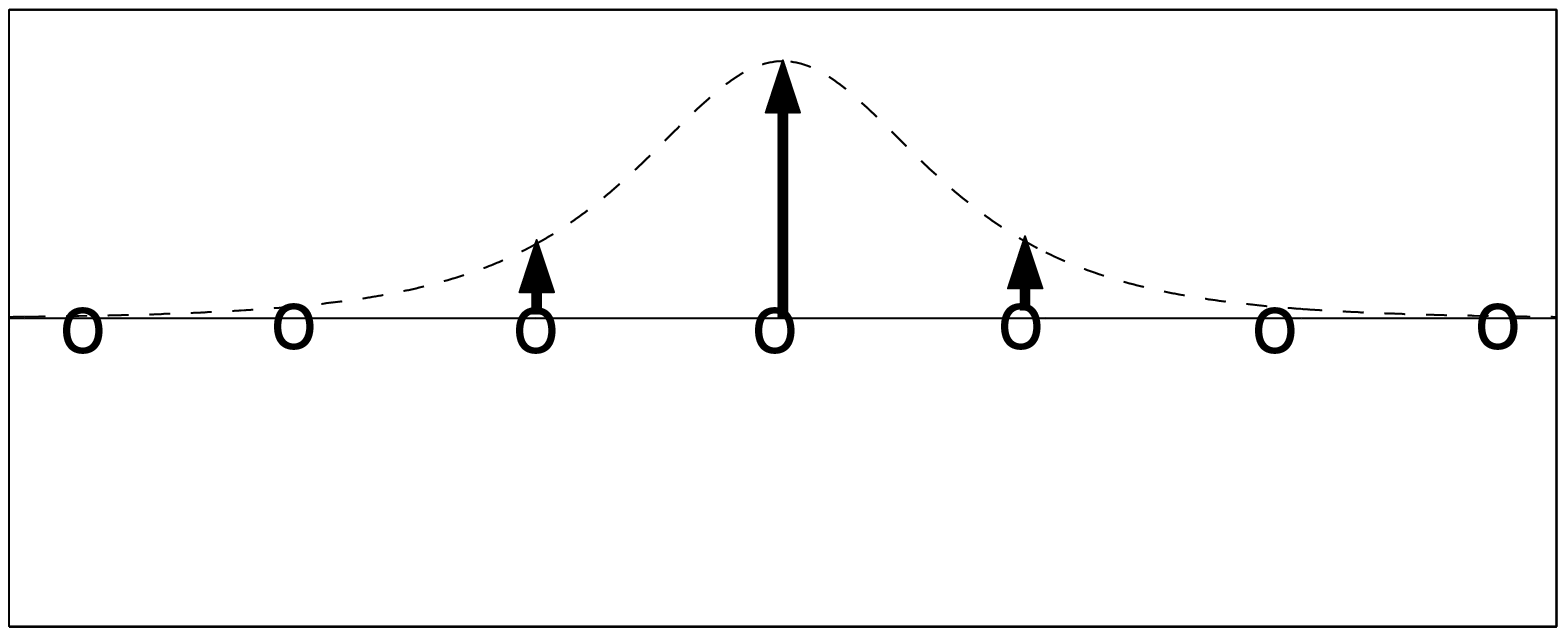}
\hskip 1.5 cm
\includegraphics{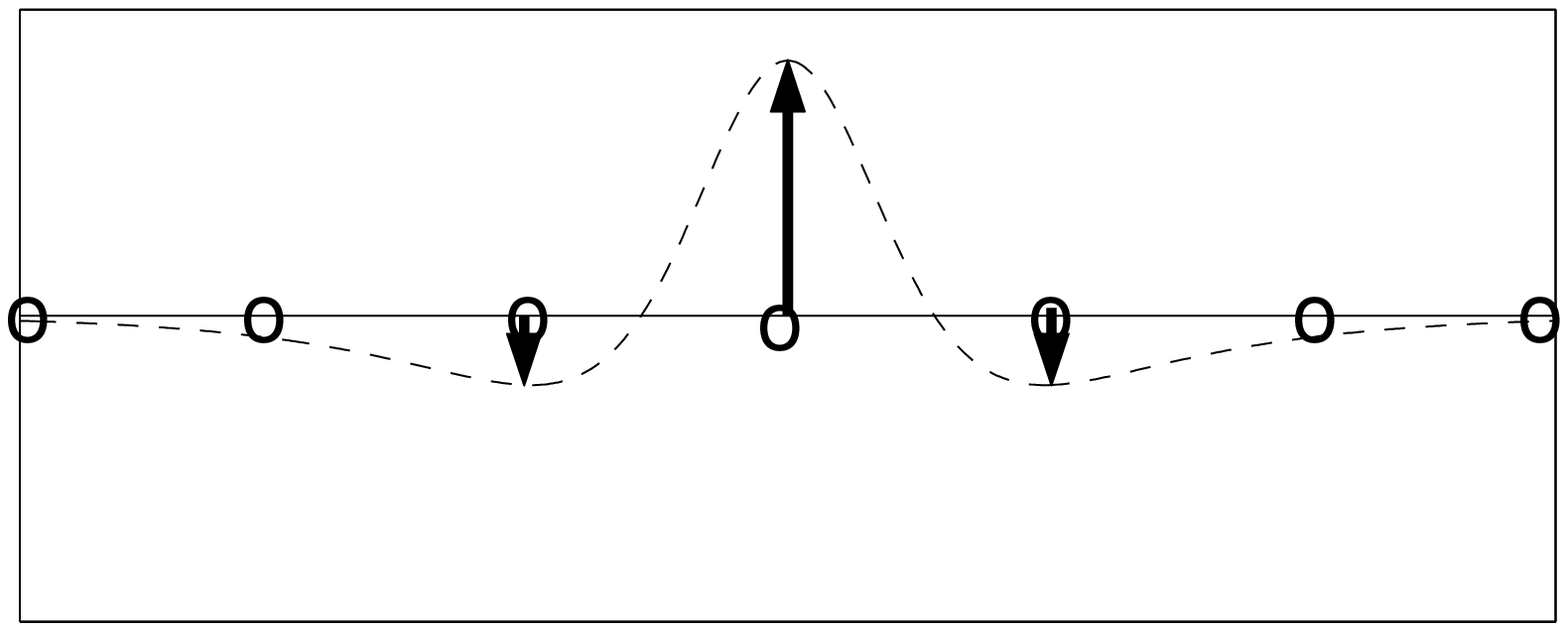}
} \caption{\small{Localized discrete breather dust lattice
excitations; the successive lattice site displacements are
depicted at maximum amplitude:
 (a) odd-parity solution; (b)
even-parity solution.}} \label{fig2}
\end{figure}

This numerical scheme is now being elaborated, and the detailed
results will be reported in an extended paper, in preparation.

\section{Breather control.}

The stability of a breather excitation may be \emph{controlled}
via external feedback, as known from one-dimensional discrete
solid chains \cite{Bountis2}. The method consists in using the
knowledge of a reference state (unstable breather), say $\delta
z_n^{(0)} = \hat z_n(t)$, e.g. obtained via an investigation of
the homoclinic orbits of the 2d map obeyed by the main Fourier
component [9], and then perturbing the evolution equation (1) by
adding a term $+ K [\hat z_n(t) - \delta z_n]$ in the right-hand
side (\textit{rhs}), in order to stabilize breathers via tuning of
the control parameter $K$. This method relies on the application
of the {\it{continuous feedback control}}  (\textit{cfc})
formalism (see the Refs. in \cite{Bountis1}). Alternatively, as
argued in \cite{Bountis1}, a more efficient scheme should instead
involve a term $+ L d [\hat z_n(t) - \delta z_n]/dt$ in the
\textit{rhs} of Eq. (1) (\emph{dissipative} \textit{cfc}), whence
the damping imposed results in a higher convergence to the desired
solution $\hat z_n(t)$. Preliminary work in this direction is
being carried out and progress will be reported later.

\section{Conclusions - discussion}

We have investigated, from first principles, the possibility of
existence of localized discrete breather-type excitations
associated with vertical dust grain motion in a dust mono-layer,
which is assumed to be one-dimensional.

 It may be noted, in concluding, that the localized
structures presented in this Letter, owe their existence to the
intrinsic lattice discreteness in combination with the
nonlinearity of the plasma sheath. Both are experimentally tunable
physical mechanisms, so our results may be investigated (and will
hopefully be verified) by appropriately designed experiments. The
experimental confirmation of their existence in dust crystals
appears as a promising field, which may open new directions e.g.
in the design of applications.

\medskip

\begin{acknowledgments}
This work was supported by the {\it{SFB591
(Sonderforschungsbereich) -- Universelles Verhalten
gleichgewichtsferner Plasmen: Heizung, Transport und
Strukturbildung}} German government Programme.

I. K. is indebted to T. Bountis (CRANS, Univ. of Patras, Greece),
S. Flach (MPIPKS, Dresden, Germany) and V. Koukouloyannis (AUTh,
Thessaloniki, Greece) for a number of elucidating discussions.
\end{acknowledgments}



\begin{thebibliography}{10}
\bibitem{psbook} P. K. Shukla and A. A. Mamun, \textit{Introduction to
Dusty Plasma Physics} (Institute of Physics, Bristol, 2002).

\bibitem{Morfill} G. E. Morfill, H. M. Thomas and M. Zuzic, in
 \textit{Advances in Dusty Plasma Physics}, Eds. P. K. Shukla,
 D. A. Mendis and T. Desai (World Scientific, Singapore, 1997) p. 99.

\bibitem{Remo} M. Remoissenet, \textit{Waves Called Solitons}
(Springer, Berlin, 1994).

\textbf{61} (10), 1443 (1973).

\bibitem{Page} S. Takeno, K. Kisoda and A. J. Sievers, Prog. Theor.
Phys. Suppl. \textbf{94}, 242 (1988); J. B. Page, Phys. Rev. B
\textbf{41}, 7835 (1990).

\bibitem{Dauxois} T. Dauxois and M. Peyrard, Phys. Rev. Lett.
\textbf{70} (25), 3935 (1993).

\bibitem{Kivshar} Yu. Kivshar, Phys. Lett. A \textbf{173} (2), 172
(1993).

\bibitem{McKay1} R. S. McKay and S. Aubry, Nonlinearity \textbf{7},
1623 (1994).

\bibitem{Flach1} S. Flach, and G. Mutschke,
Phys. Rev. E \textbf{49}, 5018 (1994).

\bibitem{Flach2} S. Flach, and C. R. Willis,
Phys. Rep. \textbf{295}, 181 (1998).

\bibitem{Chaos} See various articles in the Volume (Focus Issue):
Yu. Kivshar and S. Flach (Eds.), Chaos \textbf{13} (2), pp. 586 -
666 (2003).

\bibitem{Campbell}
D. K. Campbell, S. Flach and Yu. S. Kivshar, Physics Today,
\textbf{57} (1) (2004).

\bibitem{Jeroen} J. Bergamin, \textit{Localization in nonlinear
lattices and homoclinic dynamics}, PhD thesis, Univ. of Patras
(Faculty of Mathematics), Greece (2004).

\bibitem{Bountis1}  T. Bountis \textit{et al.}, {\it{Phys. Lett.}} A
{\textbf{268}}, 50 (2000).

\bibitem{Melandso} F. Melands\o, Phys.\ Plasmas \textbf{3}, 3890
(1996).

\bibitem{IKPKSEPJDsols}  I. Kourakis and P. K. Shukla, Eur. Phys. J. D,
\textbf{29}, 247 (2004).

\bibitem{AMS2} M. R. Amin,  G. E. Morfill and P. K. Shukla, Phys.
Plasmas {\bf 5}, 2578 (1998);
 \  Phys. Scripta {\bf 58}, 628 (1998).

\bibitem{IKPKSLDLWMI}
I. Kourakis and P. K. Shukla, Phys. Plasmas, \textbf{11}, 1384
(2004).

\bibitem{IKPKSTDLWMI}  I. Kourakis and P. K. Shukla, Phys. Plasmas,
\textbf{11}, 2322 (2004).

\bibitem{IKPKSTMDLWMI}
I. Kourakis and P. K. Shukla, Phys. Plasmas,  \textbf{11}, 3665
(2004).

\bibitem{Ivlev-Zhdanov}
A. Ivlev, S. Zhdanov, and G. Morfill, Phys. Rev. E \textbf{68},
066402 (2003).

\bibitem{Ivlev2000} A. V. Ivlev, R. S\"utterlin, V. Steinberg, M.
Zuzic and G. Morfill, Phys. Rev. Lett. \textbf{85}, 4060 (2000).

\bibitem{Misawa} T. Misawa,  N. Ohno,  K. Asano, M. Sawai, S. Takamura,
and P. K. Kaw, Phys. Rev. Lett. \textbf{86}, 1219 (2001).

\bibitem{Zafiu} C. Zafiu, A. Melzer and A. Piel, Phys. Rev. E
\textbf{63}, 066403 (2001).

\bibitem{Liu} B. Liu, K. Avinash and J. Goree, Phys. Rev. Lett.
\textbf{91}, 255003 (2003).

\bibitem{IKPKSPOPDB} I. Kourakis and P. K. Shukla,
\textit{Discrete breather modes associated with vertical dust
grain oscillations in dusty plasma crystals},
 Phys. Plasmas (in press).


\bibitem{Konopka} U. Konopka, G. E. Morfill and L. Ratke,
Phys. Rev. Lett. \textbf{84}, 891 (2000).


\bibitem{comment1} In the {\em{magnetically}} levitated dust crystal
case,
 consider the definitions in Ref.
\cite{IKPKSTMDLWMI}, upon setting $K_1 \rightarrow \alpha$, $K_2
\rightarrow \beta$ and $K_3 \rightarrow 0$ therein.

\bibitem{commentTB} Check e.g. by setting $\alpha \rightarrow
-\epsilon$ in Ref. \cite{Bountis1} and then critically comparing
the forthcoming formulae to expressions therein.

\textbf{46}, 3198 (1992).

(1992).




\bibitem{Sorasio} G. Sorasio, R. A. Fonseca, D. P. Resendes, and
P. K. Shukla, in {\it{Dust Plasma Interactions in Space}}, Nova
Publishers (N.Y, 2002), p. 37.

\bibitem{Aubry} S. Aubry, Physica D \textbf{103}, 201(1997).

\bibitem{Sepulchre} J. - A. Sepulchre and R. S. McKay, Nonlinearity
\textbf{10}, 679 (1997).

\bibitem{McKay} R. S. McKay and J. - A. Sepulchre, Physica D
\textbf{119}, 148 (1998).

\bibitem{Bountis2} T. Bountis, J. Bergamin and V. Basios, {\it{Phys.
Lett.}} A {\textbf{295}}, 115 (2002).




\end{thebibliography}
\end{document}